# Three-atom magnets with strong substrate coupling: a gateway towards non-collinear spin processing

J. Hermenau[1], J. Ibañez-Azpiroz[2], Chr. Hübner[1], A. Sonntag[1], B. Baxevanis[3], K. T. Ton[1], M. Steinbrecher[1], A. A. Khajetoorians[1,4], M. dos Santos Dias[2], S. Blügel[2], R. Wiesendanger[1], S. Lounis[2], and J. Wiebe[1]

[1] Department of Physics, Hamburg University, D-20355 Hamburg, Germany

[2] Peter Grünberg Institute and Institute for Advanced Simulation, Forschungszentrum Jülich & JARA, Jülich, Germany

[3] Leiden Institute of Physics, Leiden University, 2333 CA Leiden, The Netherlands

[4] Institute for Molecules and Materials (IMM), Radboud University, 6525 AJ Nijmegen, The Netherlands

*Corresponding author: jwiebe@physnet.uni-hamburg.de

**A cluster composed of a few magnetic atoms assembled on the surface of a nonmagnetic substrate is one suitable realization of a bit for future concepts of spin-based information technology. The prevalent approach to achieve magnetic stability of the bit is decoupling the cluster spin from substrate conduction electrons in order to suppress spin-flips destabilizing the bit. However, this route entails less flexibility in tailoring the coupling between the bits which is ultimately needed for spin-processing. Here, we show using a spin-resolved scanning tunneling microscope, that we can write, read, and store spin information for hours in clusters of only three atoms strongly coupled to a substrate featuring a cloud of non-collinearly polarized host atoms, a so called non-collinear giant moment cluster (GMC). The GMC can be driven into a Kondo screened state by simply moving one of its atoms to a different site. Owing to the exceptional atomic tunability of the non-collinear substrate mediated Dzyaloshinskii–Moriya interaction, novel concepts of spin-based information technology get within reach, as we demonstrate by a logical scheme for a four-bit register.**

Information technology is currently changing the paradigm from the separate use of the charge and spin degrees of freedom of the electrons towards a combination of both properties in so called spin-electronic elements (*1*). Using the scanning tunneling microscope as a tool, arrays of a well-defined number of atoms and shape can be assembled on a nonmagnetic substrate (*2*), and their spin state (*3*) and excitations (*4*) can be read out atom-by-atom. Recently, corresponding arrays consisting of magnetic atoms have been intensely studied as the basic constituents for future spin-based information storage (*5, 6*) and processing (*7*) schemes.

One of the main hurdles in the former respect is to achieve long spin-energy relaxation and -decoherence times (*8, 9*) of the spin states, which is usually established by two ingredients. First, a large energy barrier separating the spin states, provided by the so-called magnetic anisotropy energy (MAE) (*10*), is needed, in order to prevent the system from undergoing thermally-induced fluctuations at sufficiently low temperatures. Second, the spin states need to be protected from quantum tunneling (*11*), and from scattering by electrons (the so-called Kondo scattering (*12*)) or by phonons (*13*) from the substrate, which can shortcut the MAE barrier. To this end, an appropriate tuning of the MAE (*14*) and a *decoupling* of the spins of the array from the substrate electrons using rare earth atoms (*15, 16*) in combination with thin decoupling layers (*13, 17, 18*), superconducting (*19*), or semiconducting substrates (*20*), was usually pursued. However, with respect to spin processing, a tunable communication between the array-spins is essential. Following the decoupling layer approach, communication with stable spin states is so far restricted to dipolar (*21*) and superexchange coupling which offers limited flexibility.

A much larger flexibility can be achieved by the contrary, i.e. *strongly coupling* the array-spin to substrate electrons leading to Ruderman-Kittel-Kasuya-Yosida (RKKY) interactions (*22*), which can transfer spin information between two arrays (*7*), and can be tuned in strength, sign and even non-collinearity (*23*). In view of applications, it is therefore a formidable task to realize an array of few atoms with a long spin-energy relaxation time useable for information storage, which at the same time, is sufficiently coupled to substrate conduction electrons in order enable strong RKKY coupling to the array. Moreover, the use of heavy substrate materials would additionally feature strong spin-orbit coupling, leading to highly non-collinear Dzyaloshinskii–Moriya contributions to the RKKY interaction (*23*) which can potentially add functionality to the system.

Long-range interactions are typically promoted in a substrate material that almost fulfills the Stoner criterion for ferromagnetism, such as Rh (*24*), Pd (*25*) or Pt (*10, 22, 23, 26-29*). As a consequence, such materials display a large cloud of magnetically polarized host atoms, leading to a strongly enhanced total magnetic moment of the impurity-polarization cloud system, which was therefore referred to as a *giant moment* (GM) in the early days of research on magnetic alloys (*30*). Theses heavy substrate materials also support non-collinear magnetization states of the GM systems and Dzyaloshinskii–Moriya interactions (*23*), and are thus ideal candidates for the pursuit of a magnetically stable cluster which can be accessed by strong non-collinear RKKY interactions. With respect to information storage and processing applications, the main open questions are, thus, whether the MAE of such a non-collinear GM system can be tuned such that Kondo screening (*31*) and quantum tunneling processes (*13*) are avoided and the system is driven to support stable spin states as in a classical magnet (*32*).

## Results

### GMC fabrication

In order to realize potentially stable magnets directly coupled to a heavy substrate with strong spin-orbit coupling, we build giant moment clusters (GMCs), each consisting of three iron (Fe) atoms that are assembled into a triangular array on the (111) surface of a platinum (Pt) single crystal (Fig.1, see Methods). Each Fe atom can sit on one of the two hollow adsorption sites, face centered cubic (fcc) or hexagonally closed packed (hcp), on the lattice of the Pt surface atoms (*33*). The Fe atoms are free from adsorbed hydrogen (*34*) as indicated by their small apparent height (see Supplementary Fig.1). By lateral atom manipulation three Fe atoms are placed artificially on next nearest neighboring adsorption sites of the same type in the most compact triangular geometry (Fig.1 **a**). In this way, four different species of GMCs can be built (Fig.1**b**, see Supplementary Note 1 and Supplementary Fig.1 for the unambiguous determination of the stacking). For each of the two adsorption sites, fcc and hcp, the cluster can be either positioned such that a Pt atom is underneath the center of the trimer, referred to as *top*, or such that a hollow site is underneath the center, referred to as *hollow*. The two hollow site trimers have a larger apparent height (≈ 250 pm) in comparison to the top site trimers (≈ 200 pm). The electronic properties of all four trimer types are characterized by (inelastic) scanning tunneling spectroscopy ((I)STS, see Methods) as shown in Fig.1**c**. Their spectra show distinct excitations (arrows), which are markedly different from the excitations of isolated Fe atoms outside of trimers (*33*), indicating the emergence of a coherent electron state after the formation of the trimer.

### DFT calculations

Before we proceed with the detailed experimental investigation of these excitations, we describe spin-polarized density functional theory (DFT) calculations within the Korringa-Kohn-Rostoker Green function (KKR-GF) method (Fig.2, see Methods). Such calculations were performed in order to study the non-collinear GMC character of the trimers, which is expected from the large atomic mass and the Stoner enhancement of the substrate (*23, 29*). We find, that, for all four cluster types, the magnetic moments of the three iron atoms are coupled approximately ferromagnetic with an exchange interaction $J$ of several ten to hundred millielectron-volts (see Eq. 4 in Supplementary Note 2 for the used Hamiltonian and Supplementary Table 2 for the parameters). Moreover, there is a large cloud of more than a hundred spin-polarized Pt atoms in the vicinity of each cluster (Fig.2**a**), adding up to a total magnetic moment of the whole cluster/substrate system of $m_{\text{tot}} \sim 12 \mu_B$ for fcc hollow/top and hcp top, and of $m_{\text{tot}} \sim 13 \mu_B$ for hcp hollow (see Supplementary Table 1 for the distribution of spin and orbital moments). Closer inspection shows that, while there are considerable non-collinearities of the spins in the polarization cloud, the Fe atom spins have tilting angles of only $\theta \leq 4°$, since the Dzyaloshinskii–Moriya contribution $D_\parallel$ to the exchange is an order of magnitude smaller than $J$. The only exception is the hcp top cluster, where $\frac{D_\parallel}{J} \sim 0.5$ leads to tilting angles of up to $\theta \sim 17°$ (see Supplementary Table 2). The clusters' magnetizations have strong magnetic anisotropy energies (MAEs) $K$ on the order of one milli-electronvolt per Fe atom (Fig.2**b**,**c**). Interestingly, the polarization cloud has an RKKY-like contribution to the MAE (*27*), as revealed by its behavior as function of the number of Pt atoms considered in the collinear calculation (Fig.2**b**). For a sufficient number of Pt atoms taken into account, the calculations predict a preferred orientation of the magnetization perpendicular to the surface ("out-of-plane") for both fcc clusters, and a preferred orientation of the magnetization in the surface plane ("easy-plane") for the hcp hollow cluster. Interestingly, the MAE favors the same orientation as in the system where the constituent atom is individually adsorbed on the surface (*29*). The case of hcp top cluster is peculiar in that, although the MAE favors an easy-plane orientation by approximately 1.5 meV/adatom in the collinear calculation (see Fig. 2**b**), the strong non-collinearity emerging from the Pt substrate reverses the trend and favors instead a nearly out-of-plane orientation by ~0.1 meV/adatom in the non-collinear calculation (Fig. 2**c** and Supplementary Fig.5). The condition for the occurrence of this remarkable effect can be cast into the simple expression $\left|\frac{9 D_\parallel^2}{4J}\right| > 3K$ (see Eq. 21 in Supplementary Note 2), which shows that the effective anisotropy of the full trimer results from the competition between the non-collinear Dzyaloshinskii–Moriya contribution and the MAE. All in all, our DFT calculations reveal considerable MAEs of all four GMCs and a strong Dzyaloshinskii–Moriya–type coupling to the substrate conduction electrons which is reflected by the non-collinear substrate polarization cloud.

**Effective spin model**

In order to get a rough idea of the spin degrees of freedom of these GMCs, we map the DFT results onto an effective spin Hamiltonian, which will be later used to link the experiments to the *ab-initio* results. Considering the application of a magnetic field $B$ in the out-of-plane direction, the lowest-order approximate effective spin Hamiltonian reads $\hat{H} = g\mu_B \hat{S}_z B + \mathcal{D}\hat{S}_z^2$ (*11*). Here, $\hat{S}_z$ is the operator of the out-of-plane component of the total cluster effective spin, $g$ is the Landé g-factor, $\mu_B$ is the Bohr magneton, and $\mathcal{D}$ is the longitudinal magnetic anisotropy constant. Since the Fe atoms and their nearest neighbor Pt atoms are strongly exchange-coupled among each other (see Supplementary Table 2), the GMC is treated in this Hamiltonian as a single object with a well-defined macro spin. Neglecting the slight non-collinearity of the three ferromagnetically coupled Fe atoms plus the neighboring Pt atoms in the cluster, the DFT calculated total magnetic moments result in a mapping to a total spin with quantum number of $S = -\frac{1}{2} + \sqrt{\left(\frac{m_{\text{tot}}}{g\mu_B}\right)^2 + \frac{1}{4}} \approx 5$ or $\frac{11}{2}$ for the fcc hollow/top and hcp top GMCs, and of $S \approx 7$ or $\frac{15}{2}$ for the hcp hollow GMC. Here we considered the $g$-factors extracted from ISTS as given in Table 1. Please note, that the effective spin model is a crude approximation for the case of magnetic impurities strongly interacting with a metallic substrate, since charge fluctuations can inhibit well-defined spin quantum numbers (*34*). In the following, we choose the closest half-integer spin values ($\frac{11}{2}$ for fcc hollow/top and hcp top, and $\frac{15}{2}$ for hcp hollow) as we will see below that the hcp hollow cluster shows the Kondo effect, which strongly indicates a half integer spin ground state. However, the following discussion remains valid for the other choices of $S$ by adjusting $\mathcal{D}$ accordingly. Taking into account the positive and negative values of $\mathcal{D}$ as estimated from the DFT-derived $K$-values for the easy-plane and out-of-plane clusters (see Table 1), the resulting qualitative energy level diagrams of the spin Hamiltonian for the two cases are shown in Fig.2**d** and **e**, respectively, in zero magnetic field. We expect a degenerate $S_z = \pm\frac{1}{2}$ ground state for the hcp hollow cluster, and a degenerate $S_z = \pm\frac{11}{2}$ ground state for the fcc hollow/top and hcp top clusters. The former case is amenable to a quantum-mechanical superposition of the two $\frac{1}{2}$ spin states of the effective spin with the spin states of the substrate conduction electrons involving Kondo-correlations (*34, 35*), while the latter typically favors magnetic bi-stability, similar to a classical magnet (*5*). We therefore expect that the magnetic properties of the hcp hollow cluster will be drastically different from that of the other clusters.

**GMC spin excitations**

First, we experimentally investigated the excitations (Fig.1**c**) of all four clusters by (I)STS as a function of magnetic field $B$, and for the hcp hollow GMC also as a function of temperature $T$. Indeed, for the hcp hollow cluster (Figs.3**a,c**), we find a Fano-resonance right at the Fermi energy $E_\mathrm{F}$ (i.e. at $V = 0~\mathrm{meV}$) that splits and broadens as a function of $B$ and $T$, respectively. These are clear hallmarks for an emergent Kondo screening of the spin of this GMC (*34-38*). The asymmetric lineshape of the Fano resonance is caused by the interference of a direct tunneling into the Kondo resonance, which arises as a consequence of the zero-energy spin flip scattering (*39*), and other tunneling channels (*40*). We can adequately fit the spectra to a sum of two Fano functions using the same form factor $q$ and full width at half maximum $\Gamma$, but allowing for different energetic positions $E_i$ of the Kondo resonances (lines in Figs.3**a**,**c**, see the used Fano functions and parameters in Supplementary Note 3, Supplementary Tables 3,4, and Supplementary Fig.6). The extracted values are used to quantify the field-induced splitting and temperature-driven broadening, as shown in Figs.3**b**,**d**. Indeed, the resonance shifts linearly with $B$ (Fig.3**b**) with a *g*-factor close to 2 ($g_\mathrm{hcp\ hollow} \sim 1.7 \pm 0.2$), which we expect from the Kondo screening of the spin $\frac{1}{2}$ ground state of the easy-plane anisotropic hcp hollow cluster (Fig.2**d**) (*34, 35*). The dependence of $\Gamma$ on temperature (Fig.3**d**) nicely fits to a power law and to numerical renormalization group calculations for a spin-1/2 impurity (*41*) (see Supplementary Note 3) using a Kondo temperature of $T_K \sim 4.5~\mathrm{K}$. This value is one to two orders of magnitude smaller as compared to usual Kondo temperatures of single impurities on noble metal surfaces (*37*), but has a similar size as that of atoms which are decoupled from the conduction electrons by thin insulators (*35*) or that of Fe-hydrogen complexes on Pt(111) (*34*). Our experimental findings thus prove Kondo correlations of the hcp hollow GMC with the substrate conduction electrons.

The magnetic-field evolution of the excitations of the fcc hollow/top and hcp top clusters behave drastically different (Fig.1**c**, Supplementary Note 4, Supplementary Figs.7,8). The spectra on these GMCs reveal a pair of features at energies symmetric to $E_\mathrm{F}$ which linearly shift away from $E_\mathrm{F}$ with increasing magnetic field. This behavior is very reminiscent of systems with an out-of plane MAE (*29*) which is excited from the ground state into the first excited state over the zero field energy splitting of $\Delta_{01} = \mathcal{D}(2S - 1)$ (Fig.2**e**) that is linearly increasing by Zeeman energy. Note, however, that the shapes of the features deviate from simple steps, which probably indicates the limitation of the effective spin model (see Supplementary Note 4). By measuring the energetic positions of the excitations as a function of $B$ (Supplementary Figs.7,8 and Supplementary Table 5) we extract the corresponding longitudinal anisotropy constants $\mathcal{D}$ and *g*-factors close to 2 which are given in Table 1. The anisotropy constants for the two fcc and the hcp top clusters fit reasonably well with the constants estimated from the DFT calculations (see Table 1), substantiating the out-of-plane character of these GMCs.

**GMC spin dynamics**

Due to the out-of-plane MAE of the fcc top/hollow and hcp top GMCs, we expect a bi-stable switching between the $S_z = \pm\frac{11}{2}$ ground states, which is observable if there are not too frequent tunneling electrons with an energy larger than $\Delta_{01}$ which will drive the system across the barrier between these two states by sequential excitations (Fig.2**e**) (*5*). On the other hand, the emergent Kondo behavior of the hcp hollow GMC should lead to a quantum mechanical superposition of the $S_z = \pm\frac{1}{2}$ states of the effective spin with the spin states of substrate conduction electrons such that there is no observable switching of the magnetization. Indeed, using Cr coated tungsten tips which are sensitive to the out-of-plane component of the GMC magnetization (Methods and Supplementary Fig.9), we observe striped patterns in constant-current images taken on fcc top/hollow and hcp top clusters (shown exemplarily for the fcc top cluster in Fig.4**a**), but not on the hcp hollow GMC. Such stripe patterns result from a random two-state telegraph noise (*5, 32*), which is visible if the tip is positioned on-top of those clusters and the height is measured as a function of time (Fig.4**b**). The telegraph signal can be unambiguously assigned to a statistical switching of the GMC between two spin states (0) and (1) of opposing out-of-plane magnetizations (Fig.4**d**) by verifying that the asymmetry of the time-averaged occupational lifetimes $\bar{\tau}_0$ and $\bar{\tau}_1$ of the two states, defined by $\mathcal{A} = \frac{\bar{\tau}_1}{\bar{\tau}_0 + \bar{\tau}_1}$, reverses by either reversing the out-of-plane oriented magnetic field $B$, or by reversing the out-of-plane spin-polarized tunneling current (bias polarity), which both drive the spin into one of the two states (*5, 32*). We prove, that this is indeed the case for the fcc top/hollow and hcp top GMCs as shown in Figs.4**b**,**c**. Similarly, full $B$-dependent loops of the asymmetry for two bias polarities recorded for all three clusters (Fig.4**e**) show all the characteristics of a system randomly switching between two out-of-plane spin states and driven by the spin-pumping effect of the spin-polarized tunneling electrons (*5*). Note, that there is a considerable shift of these asymmetry curves towards negative magnetic fields which indicates an effective stray or exchange field of the tip (*3, 42*) on the order of $-0.1$ T. Moreover, the asymmetry curve recorded on the fcc hollow GMC is apparently inverted with respect to the vertical axis, which might be related with an inverted vacuum spin-polarization of this cluster as compared to the other two clusters, as sensed by the tip apex.

We consequently identify the two spin states (0) and (1) observed in the telegraph noise with the two ground states $S_z = +\frac{11}{2}$ and $S_z = -\frac{11}{2}$ of the fcc top/hollow and hcp top GMCs (Fig.2**e**). In order to study the dynamics in these spin states in detail, we measure the telegraph noise as a function of bias voltage, tunneling current, and temperature (Figs.4**f**-**h**). The temperature dependence of the average lifetime $\bar{\tau} = \left(\frac{1}{\tau_0} + \frac{1}{\tau_1}\right)^{-1}$ (Fig.4**f**) is slightly different for all three clusters and on the order of a few tens of seconds at $T = 300 \text{ mK}$. It reveals an exponential decay with increasing temperature which can be assigned to a quasi-classical Arrhenius-like behavior that has been observed for larger scale Fe islands (*32*) and non-$C_{3v}$-symmetric clusters at larger temperatures (*5*). The absence of any shoulder in the temperature dependence down to our lowest measurement temperature indicates, that transversal anisotropies which could induce quantum tunneling processes, are negligible here (*5*). Furthermore, as shown in Fig.4**g**, $\bar{\tau}$ increases over more than two decades if the bias voltage approaches the excitation thresholds $\frac{\Delta_{01}}{e}$ determined from ISTS (Supplementary Table 5), and finally gets too long for a statistically relevant measurement. This behavior is indeed expected, considering that, without quantum tunneling through the barrier, the spin state of the GMC can only overcome the barrier by sequential excitation processes, which need either tunneling or substrate electrons with energies larger than $\Delta_{01}$. Indeed, we can consistently reproduce all measurements by using a master equation approach (see Methods) in conjuction with the effective spin Hamiltonian given above, with $S = \frac{11}{2}$, the *g*-factors extracted from ISTS (see Table 1) and fitting the longitudinal magnetic anisotropy parameter $\mathcal{D}$ (lines in Figs.4**e-g**). The resulting magnetic anisotropies given in Table 1 agree with the parameters determined from ISTS remarkably well, giving a final support for the quasi-classical spin behaviour of the fcc top/hollow and hcp top GMCs, in stark contrast to the Kondo character of the hcp hollow GMC.

**GMCs as two-bit and four-bit registers**

As suggested by Fig.4**g**, a spin-polarized tunneling current of sufficiently small bias below the excitation threshold $\frac{\Delta_{01}}{e}$ could be used to "read" the quasi-classical spin state of the fcc top/hollow and hcp top GMCs with negligible disturbance. On the other hand, a current biased above this threshold could be used to switch the spin state and thereby "write" information. This is verified by building a two-bit register from two of the magnetically most stable fcc top GMCs located next to each other in a distance of only $2.5 \text{ nm}$ (Fig.5**a-d**). Indeed, we were able to write the state (0) or (1) into the desired GMC by application of a spin-polarized tunneling current driven by a voltage-ramp crossing $\frac{\Delta_{01}}{e}$ that was switched off when the GMC switched from (0) to (1), or vice versa, at wish, without changing the state of the neighbouring GMC. The consecutive read out of the four possible states (01), (11), (10), and (00) of the two-bit register written like this is shown in Fig.5**a-d**. Furthermore, we demonstrated, that such prepared states are stable in time by reading the spin state of one of the CMCs prepared in state (0), while the other was in state (1), over time. Figure 5**e** shows that the state is stable for at least 10 hours, after which we had to stop the measurement due to restrictions imposed by the experimental facility.

Since the GMCs are strongly coupled to a heavy metallic substrate that was experimentally shown to mediate Dzyaloshinskii–Moriya interactions tunable by the distance (*23*), the realization of a two-bit register of spins, demonstrated here, offers further perspectives for spin-based information processing schemes. One possible scheme made from the two-bit register in Fig.5**a-d** and an additional hcp-hollow cluster (or alternatively an Fe atom on hcp adsorption site which has easy-plane anisotropy (*34*)) in close to perpendicular geometry is shown in Fig.5**f**. The hcp-hollow cluster should be assembled at a distance where the substrate mediated interaction is strong enough to quench the Kondo screening, permitted by its small Kondo temperature ($k_B T_K \sim 0.4$ meV). Since the Dzyaloshinskii–Moriya vector $\boldsymbol{D}$ has a strong contribution in the plane and perpendicular to the dotted line (see Fig.5**f**) (*23*), the spin of the hcp hollow cluster (or hcp atom) can be forced into the four different orientations shown in Fig.5**f-i** depending on the four possible states of the two-bit register.

## Discussion

We finally compare the magnetic stability of the three-atomic fcc top GMC investigated here with the system of five Fe atoms constructed on the (111) surface of copper (*5*). As shown in Fig.4**h**, under writing conditions, a number of about $\mathcal{N} = \bar{\tau} \cdot I = 10^8$ tunneling electrons (at $0.75$ nA current and $5$ mV bias, slightly depending on the current) is needed for a single switching event of the fcc top GMC. We investigated other fcc top clusters using different tips and found that $\mathcal{N}$ can be as large as $10^9$ ($0.75$ nA, $10$ mV), where the variation between clusters results from small variations in the MAE due to electronic inhomogeneity of the substrate (*5*). This number is about five to ten times smaller than that needed to switch $Fe_5$ on Cu(111) at comparable tunneling parameters. The increased sensitivity under writing conditions might be explained by the smaller number of atoms in the GMC investigated here, leading to a smaller spin and consequently to a smaller number of states separating the two spin states (0) and (1). Fully understanding why $Fe_3$ on Pt(111) still is a magnetically stable GMC requires a detailed analysis of diverse effects, such as symmetry protection (*14*) or quantum spin-fluctuations. For instance, the higher symmetry of the $Fe_3$ on Pt(111) GMC leeds to negligible transversal anisotropies efficiently reducing quantum mechanical tunneling of the spin through the MAE barrier, which is not the case for the less symmetric $Fe_5$ on Cu(111) system. Moreover, quantum spin-fluctuations can be suppressed if the hybridization with the substrate is less effective (*43*), which appears to be the case for Pt(111) in comparison to Cu(111), according to our preliminary results. These effects are likely to be the reason for the fact, that the $Fe_3$ on Pt(111) GMC can still serve as a stable magnetic bit under reading conditions, while it needs less electrons to switch its state under writing conditions.

In summary, we experimentally demonstrated that a GMC of only three Fe atoms strongly coupled to a heavy-element substrate constitutes a stable quasi-classical magnet suitable for storing information for hours. Due to the strong coupling of the magnetic bit to a substrate that features non-collinear RKKY-interactions (*23*), the spin-information can be processed with large variability, as we demonstrate by the scheme of a four-bit register. Moreover, the chosen substrate platinum is a prototypical strong spin-orbit coupling material widely used in the research on spintronics technology concepts. Therefore, the few-atom magnet we have realized here is an ideal model system to study the down-scaling of such concepts, as, e.g., the application of spin-orbit torques for writing information (*44, 45*), to the limit of single atoms. The presented atomic-scale magnetic information storage with tunable interactions therefore offers huge potential flexibility for the study of future nanospintronics components.

## Methods

### Experimental procedures

All measurements have been performed under ultra-high vacuum conditions in a home-built low-temperature scanning tunneling microscope facility were the magnetic field $B$ is applied perpendicular to the sample surface (*46*). The Pt(111) single crystal was cleaned in situ by argon ion sputtering and annealing cycles and a final flash as described in ref. (*29*). While the sample cooled down to room temperature after the flash, a fraction of a monolayer Co was deposited from an e-beam heated rod resulting in the decoration of the Pt step edges and terraces with Co stripes and islands, respectively, of one atomic layer height (*3, 28*) (see Supplementary Fig.9). After the sample was cooled to $T \sim 4$ K, ~1% of a monolayer Fe was co-deposited from an e-beam heated rod. During deposition, the sample temperature did not exceed $T \sim 10$ K resulting in a statistical distribution of single Fe atoms on the Pt terraces (Fig. 1a and Supplementary Fig.9).

Constant-current images were recorded at a tunnelling current $I$ with a bias voltage $V$ applied to the sample. The Fe clusters have been formed by lateral manipulation of Fe atoms using a current of $I = 25 - 35$ nA with voltages of $V = 0.9 - 1.3$ mV (*2, 7*). For spin-resolved scanning tunneling microscopy (SPSTM), we coated flashed tungsten tips with about 50 monolayers of Cr (*33*). Spin contrast with a sensitivity to the out-of-plane component of the sample magnetization was then achieved by picking up Fe atoms until a strong magnetic contrast was observed on the remanently out-of-plane magnetized Co stripes or islands (*3, 28*) (Supplementary Fig.9). For the investigation of the spin excitations by spin-averaged ISTS we dipped the tip into the Pt substrate until the telegraph signal on the out-of-plane clusters vanished and the spectrum taken on the Pt was feature-less within the voltage range used for ISTS. Then, ISTS was performed by positioning the tip on top of the GMC, adding a modulation voltage $V_{\text{mod}} = 80\ \mu V$ (r.m.s.) of frequency $f_{\text{mod}} = 4142\ \text{Hz}$ to $V$, stabilizing the tip at $I_{\text{stab}} = 2\ \text{nA}$ and $V_{\text{stab}} = 5\ \text{mV}$, switching the feedback off, ramping the bias voltage and recording the $\frac{dI}{dV}$ signal using a lock-in amplifier. For the investigation of the spin dynamics of the GMCs by SPSTM, we moved an out-of-plane spin-sensitive tip on top of the cluster, waited until the scanner creep was negligible, and recorded the tip height $z$ as a function of time in constant-current mode (*5*).

### *Ab initio* DFT calculations

For the *ab initio* DFT calculations we have used the Korringa-Kohn-Rostoker Green function (KKR-GF) method including spin-orbit coupling within the local-spin-density approximation (*47, 48*). The Pt(111) hollow and top clusters have been modeled by a slab containing 24 and 64 Pt atoms, respectively, in both cases augmented by two vacuum regions. The relaxed distances of the Fe trimers towards the surface have been calculated by means of the QUANTUM-ESPRESSO package (*49*). The MAE and the non-collinear energy landscape of the four types of trimers have been evaluated by band energy differences following the magnetic force theorem (*50*), while the magnetic interactions among the Fe atoms have been computed using the generalized Liechtenstein formula (*51-53*) and a fine mesh of 200x200x1 *k* points. For further details, we refer the reader to the Supplementary Note 2.

### Master equation model

In order to simulate the GMC dynamics, we used the master equation model in conjunction with the effective spin Hamiltonian as described in ref. (*5*). Additionally, we took into account a bias voltage dependent change in the ratio between the coupling of the cluster to the tip, $v_{\text{tip}}$, and to the surface, $v_{\text{surface}}$, which we observed for the clusters investigated here. To this end, we assumed an exponential dependence on the bias voltage $V$ via $\left(\frac{v_{\text{tip}}}{v_{\text{surface}}}\right)_V = \left(\frac{v_{\text{tip}}}{v_{\text{surface}}}\right)_{V_T} \times e^{-\kappa \cdot (V-V_T)}$, where $\left(\frac{v_{\text{tip}}}{v_{\text{surface}}}\right)_{V_T}$ is the ratio determined from the fit of the temperature dependence of the lifetime measured at voltage $V_T$ (Fig.4f). The tip retraction parameter $\kappa$ then follows from the fitting of the voltage dependence of the lifetime (Fig.4g), resulting in $\left(\frac{v_{\text{tip}}}{v_{\text{surface}}}\right)_{V_T} = 0.085$ and $\kappa = 0.15$ for hcp top, $\left(\frac{v_{\text{tip}}}{v_{\text{surface}}}\right)_{V_T} = 0.095$ and $\kappa = 0.1$ for fcc hollow, and $\left(\frac{v_{\text{tip}}}{v_{\text{surface}}}\right)_{V_T} = 0.07$ and $\kappa = 0.08$ for the fcc top cluster. Since $\frac{v_{\text{tip}}}{v_{\text{surface}}}$ additionally depends on the current, the fitted voltage dependence slightly deviates from the experimental data for the hcp top cluster. $\frac{v_{\text{tip}}}{v_{\text{surface}}}$ varies between 0.01 and 0.022 for the field dependent asymmetry fits (Fig.4e).

**Data availability**

The authors declare that the main data supporting the findings of this study are available within the article and its Supplementary Information files. Extra data are available from the corresponding author upon reasonable request.

## Acknowledgements

We thank Kirsten von Bergmann for stimulating dicussions about non-collinear spin-processing schemes, Markus Ternes for valuable suggestions concerning the fitting of the Kondo resonances, and Roman Kovacic for technical help and discussions. J.H., A.S., M.S., R.W., and J.W. acknowledge funding from the SFB668 and the GrK1286 of the DFG. A.A.K. acknowledges funding from the Emmy Noether Program of the DFG, and the Dutch funding agencies FOM, which is part of NWO, and NWO (VIDI program). J.I.-A., M.dSD.,and S.L. acknowledge funding from Helmholtz Gemeinschaft Deutscher-Young Investigators Group Program No. VH-NG-717 (Functional Nanoscale Structure and Probe Simulation Laboratory), the Impuls und Vernetzungsfonds der Helmholtz-Gemeinschaft Postdoc Programme, and the ERC consolidator grant (DYNASORE).

## Author contributions

J.H., A.S., and J.W. designed the experiments. A.A.K. and M.S. helped with the design of the initial experiments. J.H., A.S., and K.T.T. carried out the measurements. J.H. and A.S. did the analysis of the experimental data. J.I.-A., M.dSD., and S.L. performed and analysed the DFT calculations. Chr.H. and B.B. did the master equation modelling. J.H. and J.W. wrote the manuscript, to which all authors contributed via discussions and corrections.

## Tables

| GMC species | DFT $\mathcal{D}$ (meV) | spin excitations (ISTS) | | spin-dynamics (SPSTM) | |
|---|---|---|---|---|---|
| | | $g$ | $\mathcal{D}$ | $g$ | $\mathcal{D}$ |
| hcp hollow | $+0.035$ | $1.69 \pm 0.22$ | - | - | - |
| hcp top | $-0.01$ | $2.18 \pm 0.08$ | $-0.05 \pm 0.01$ | 2.2 | $-0.025$ |
| fcc hollow | $-0.25$ | $1.86 \pm 0.06$ | $-0.07 \pm 0.01$ | 1.92 | $-0.05$ |
| fcc top | $-0.03$ | $2.12 \pm 0.06$ | $-0.09 \pm 0.02$ | 2.06 | $-0.05$ |

**Table 1 | Determined values of the longitudinal anisotropy constant $\mathcal{D}$ and $g$-factors.** The values of $\mathcal{D}$ expected from the DFT calculations were estimated from the calculated $K$ (including non-collinear effects) via $\mathcal{D} = \frac{K}{S^2 - \frac{1}{4}}$ assuming $S = \frac{11}{2}$ for hcp top, fcc hollow and top, and $S = \frac{15}{2}$ for hcp hollow, which are determined from the calculated magnetic moments together with the $g$-values used in the effective spin model. The values from the spin excitations are determined using the measured zero-field spin-excitation energies $\Delta_{01}$ via $\mathcal{D} = \frac{\Delta_{01}}{2S-1}$, and the shifts of the excitations with $B$ via $g = \frac{1}{\mu_B} \frac{d\Delta_{01}}{dB}$ (Supplementary Note 4, Supplementary Figs.7,8 and Supplementary Table 5). The $\mathcal{D}$ values determined from the spin-dynamics have been obtained by fitting the master equation model in conjuction with the effective spin Hamiltonian, using $S = \frac{11}{2}$ and the given $g$-factors, to the experimental data.

# Figures

**Figure 1 | Fabrication of the GMCs. a**, Constant-current images taken during the fabrication of an $Fe_3$ cluster from three individual Fe atoms on Pt(111) (image sizes $5 \times 5 \text{ nm}^2$, $V = -5 \text{ mV}$, $I = 500 \text{ pA}$). Left: the arrow indicates the first manipulation step of one of the individual Fe atoms; Middle: a dimer has been formed, the arrow indicates the second manipulation step; Right: the $Fe_3$ cluster has been formed. **b**, Constant-current images of the four possible $Fe_3$ cluster geometries (image sizes: $2 \times 2 \text{ nm}^2$, $V = 5 \text{ mV}$, $I = 500 \text{ pA}$). The open circles indicate the underlying lattice of surface Pt atoms and the black spheres the positions of the three Fe atoms within each cluster. A height profile across the center of each of the individual clusters is indicated in the middle. **c**, ISTS taken on the center of each cluster ($B = 0 \text{ T}$, $T = 0.3 \text{ K}$). The step-like features in each spectrum originating from the excitations of the clusters are indicated by arrows.

**Figure 2 | Magnetic anisotropy and energy level schematics of the GMCs. a**, Model of the hcp-top $Fe_3$ cluster on the Pt surface resulting from the *ab-initio* (KKR) calculations indicating the strong non-collinearity of the magnetic moments (arrows) in the cluster and polarization (color) in the Pt. The inset shows the definition of the non-collinearity angle $\theta$. **b**, Collinear *ab-initio* calculated MAE of all four cluster geometries as a function of the number of considered Pt atoms. **c**, DFT non-collinear band-energy calculations (energy per Fe atom) of the hcp top cluster as a function of the non-collinearity angles. Squares and stars depict the energy evolution of nearly out-of-plane and in-plane spin configurations, respectively (see Supplementary Figs. 3 and 4). The origin of the energy axis has been set with respect to the band-energy of the collinear ($\theta = 0$) out-of-plane configuration. The horizontal solid and dashed lines respectively denote the minimum band-energy of the nearly out-of-plane and in-plane configurations, whose difference defines the effective anisotropy $K_{\text{eff}}$. **d,e**, Schematic diagrams of energy levels in zero magnetic field for the easy-plane case of the hcp hollow cluster (d, $\mathcal{D} > 0$) and the out-of-plane case of the fcc top, fcc hollow, and hcp top clusters (e, $\mathcal{D} < 0$), illustrating the effective spin model used for the simulation of the magnetization dynamics. $\Delta_{01}$ in e denotes the splitting between the ground and first excited states.

**Figure 3 | Kondo screening of the hcp-hollow GMC. a**, Colored lines are spectra taken at the indicated magnetic fields on the hcp-hollow cluster ($T = 0.3 \text{ K}$). The black lines are fits to the sum of two Fano-functions (values of the full width at half maximum $\Gamma \approx 0.5 \text{ meV}$ and of the form factor $q$, see Supplementary Note 3, Supplementary Table 3, and Supplementary Fig.6). **b**, Symbols are the magnetic field dependent energetic positions of the Kondo resonances extracted from the fitted Fano-functions in a. The solid lines are linear fits through $(E_i, B) = (0,0)$ resulting in the $g$-factors of 1.46 and 1.91 for the positive and negative energy side, respectively. **c**, Measured temperature dependence of the Kondo resonance (colored lines, $B = 0 \text{ T}$) together with fitted single Fano-functions (black lines, $E_i \approx 0 \text{ meV}$, values of $q$, see Supplementary Table 4). **d**, Temperature dependency of the width $\Gamma$ of the Kondo resonance extracted from the fitted Fano-functions in c, together with a fit to a power law (grey line) and to numerical renormalization group calculations for a spin-1/2 impurity in the strong coupling regime (grey dots, taken from (*41*)).

**Figure 4 | Magnetization switching of hcp-top, fcc-hollow, and fcc-top GMCs. a**, Spin-resolved constant-current image of a fcc-top cluster taken with an STM tip sensitive to the out-of-plane component of the cluster magnetization, showing switching of the apparent height due to magnetization switching ($3 \times 3$ nm$^2$, $V = 2$ mV, $I = 200$ pA, $B = 0$ T, $T = 0.3$ K). **b**, Spin-dependent telegraph signal measured at different magnetic fields and bias polarities as indicated on a fcc-top cluster ($I = 500$ pA, $T = 0.3$ K). Two lifetimes $\tau_0$ and $\tau_1$ for the two magnetization states are indicated. **c**, Voltage polarity- and magnetic field-dependent histograms (scale from 0 to 1) of the state-dependent lifetimes $\tau_0$ and $\tau_1$ illustrating favorability of the state 1 for positive field or positive bias (and 0 for negative field or negative bias). **d**, Sketch of the two magnetization states down (0) and up (1) comprised of the magnetization of the Fe cluster and a cloud of polarized Pt substrate atoms. **e**, Symbols indicate measured magnetic field dependent asymmetries $\mathcal{A}$ ($T = 0.3$ K, $I = 500$ pA) of hcp-top ($|V| = 1.5$ mV), fcc-hollow ($|V| = 2$ mV) and fcc-top clusters ($|V| = 5$ mV). **f**, Measured temperature dependence of $\bar{\tau}$ at $B = 0$ T (hcp top: $V = -0.7$ mV, $I = 375$ pA; fcc hollow: $V = -1.25$ mV, $I = 750$ pA; fcc top: $V = -2$ mV, $I = 750$ pA). **g**, Measured voltage dependence of $\bar{\tau}$ ($I = 750$ pA, $B = 0$ T, $T = 0.3$ K). Solid lines in e, f, and g are the corresponding model calculations using an effective spin of $S = 11/2$ with parameters of the magnetic anisotropy constants $\mathcal{D}$, the $g$-factors, and the tunnel couplings as given in Table 1 and the Methods section. Dashed lines in g indicate the energies $\Delta_{01}$ of the first excited state for each cluster type determined from ISTS (see Supplementary Table 5). **h**, Average numbers of tunneling electrons $\mathcal{N} = \bar{\tau} \times I$ needed for a single switching event as a function of the current ($B = 0$ T, $T = 0.3$ K, $V = 5$ mV). The colors in e to h indicate the different cluster types according to the colors given in a.

**Figure 5 | Two-bit and four-bit registers based on GMCs. a** to **d**, Spin-resolved constant-current images of two fcc-top clusters in the four possible spin states (01), (11), (10), and (00), where 0 and 1 correspond to downwards and upwards pointing magnetization, respectively (imaging parameters: $V = -1$ mV, $I = 500$ pA, $B = 0$ T, $T = 0.3$ K). The Fe atom in the back serves as a marker for the apparent height. Between the images, the tip was positioned on top of the cluster whose state was intended to be changed, the bias was slowly increased until the state switched, and then quickly decreased to the imaging parameter. **e**, The long-term stability of one of the two magnetic bits is shown by measuring the height of the cluster (orange trace) in the state 0 over more than 10 hours ($V = -0.7$ mV, $I = 50$ pA, $B = 0$ T, $T = 0.3$ K), without a single switch into the state 1, whose height reference is given by the red dotted line which was determined from the magnetic contrast in images a to d. **f** to **i**, Top view schematic diagram of a possible register with four different realized spin states on a hcp hollow cluster. The two upper circles in each panel illustrate two fcc top clusters prepared in magnetization states up (dot) or down (cross). The lower circle in each panel signifies a hcp hollow cluster (or a hcp-atom), which is forced into one of the four spin states indicated by the thick arrow, as dictated by the Dzyaloshinskii–Moriya interaction $\boldsymbol{D}$ (orientation indicated by the two thin arrows) to the two neighboring clusters.

**Figure 1**

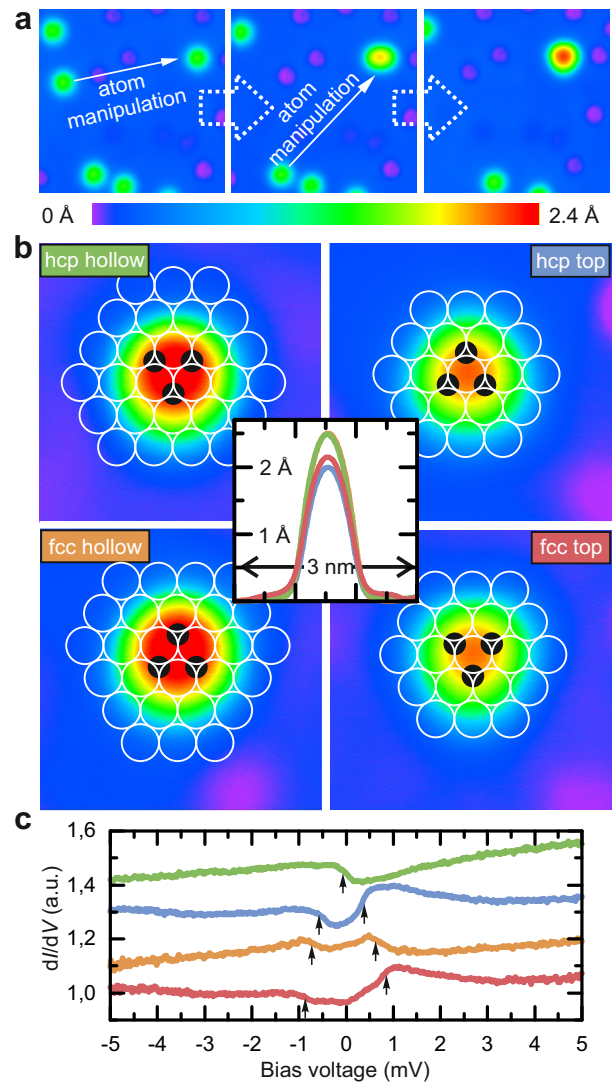

**Figure 2**

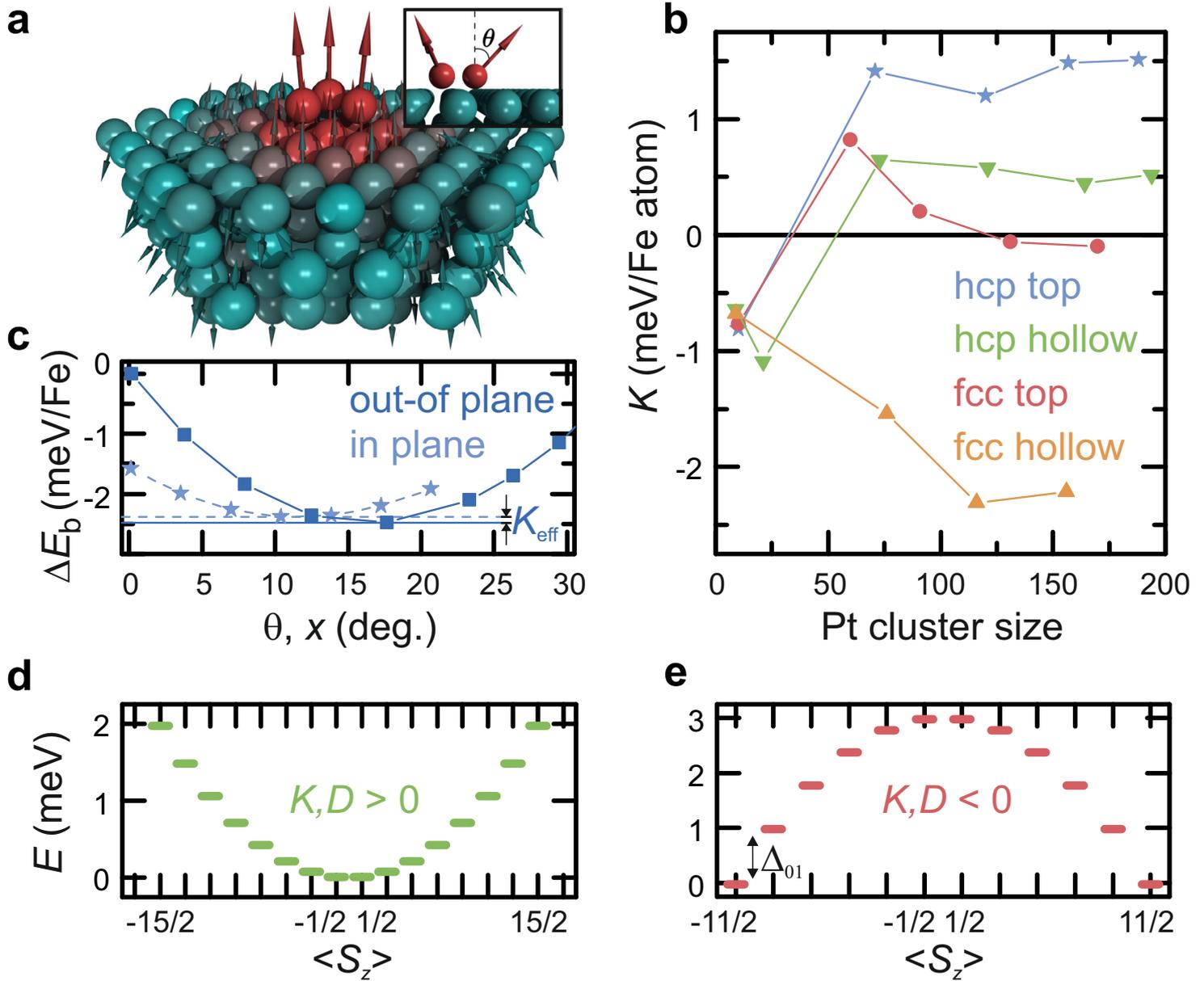

**Figure 3**

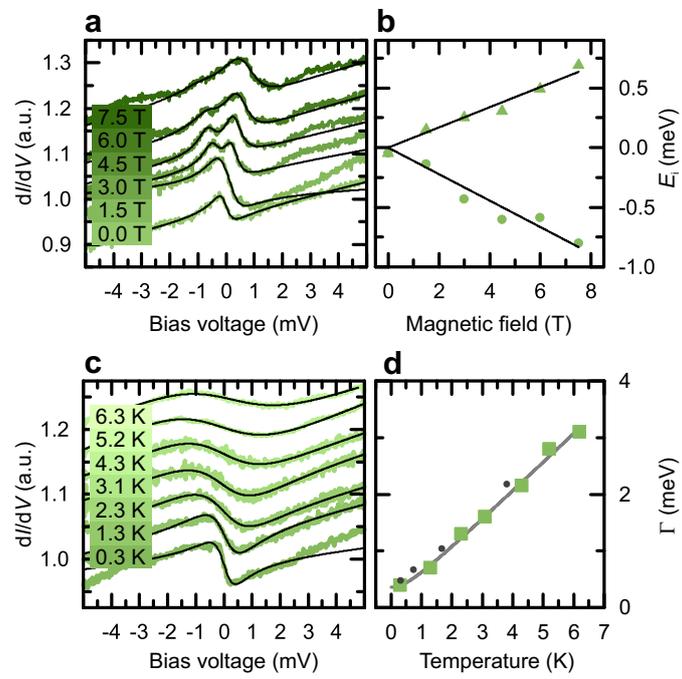

# Figure 4

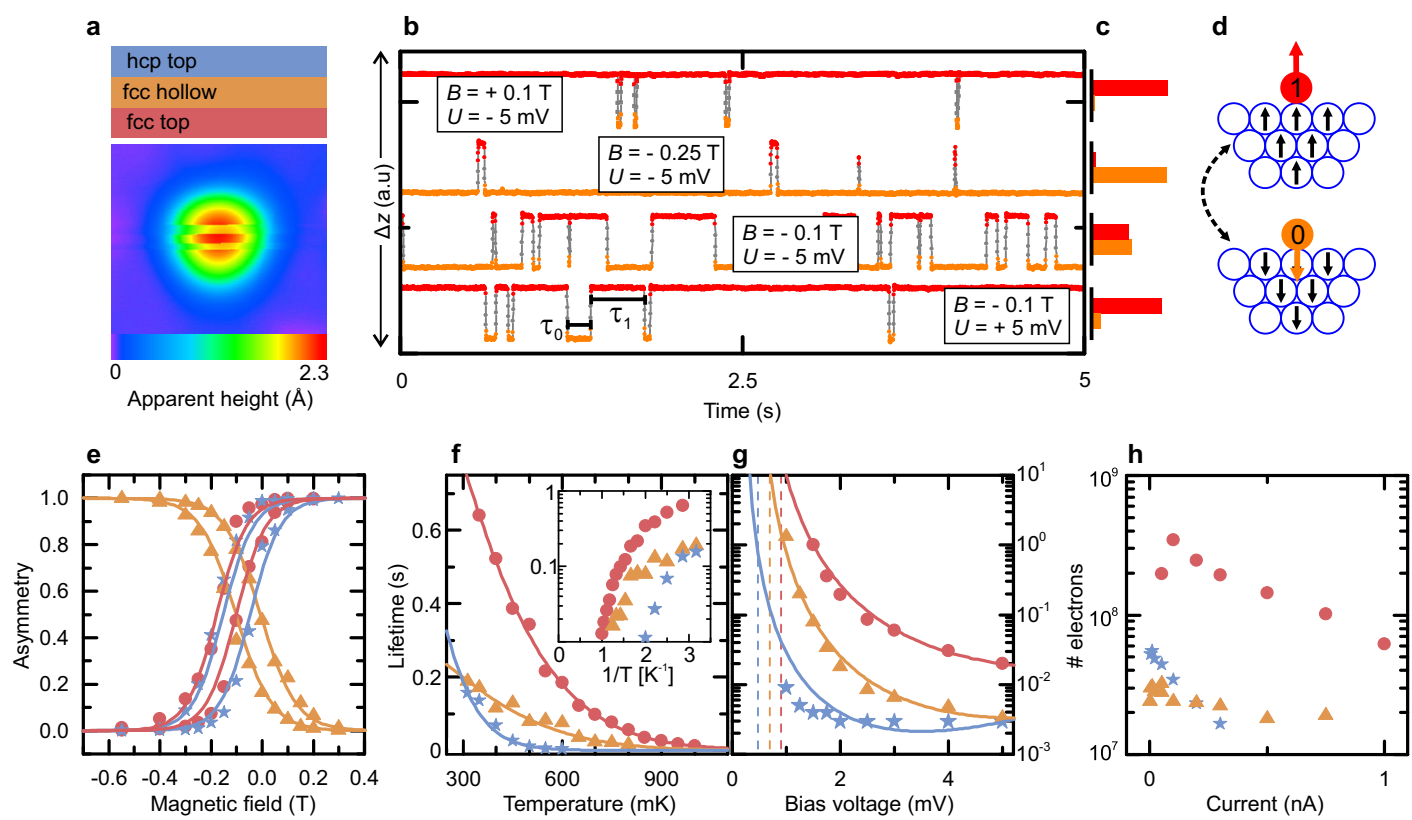

**Figure 5**

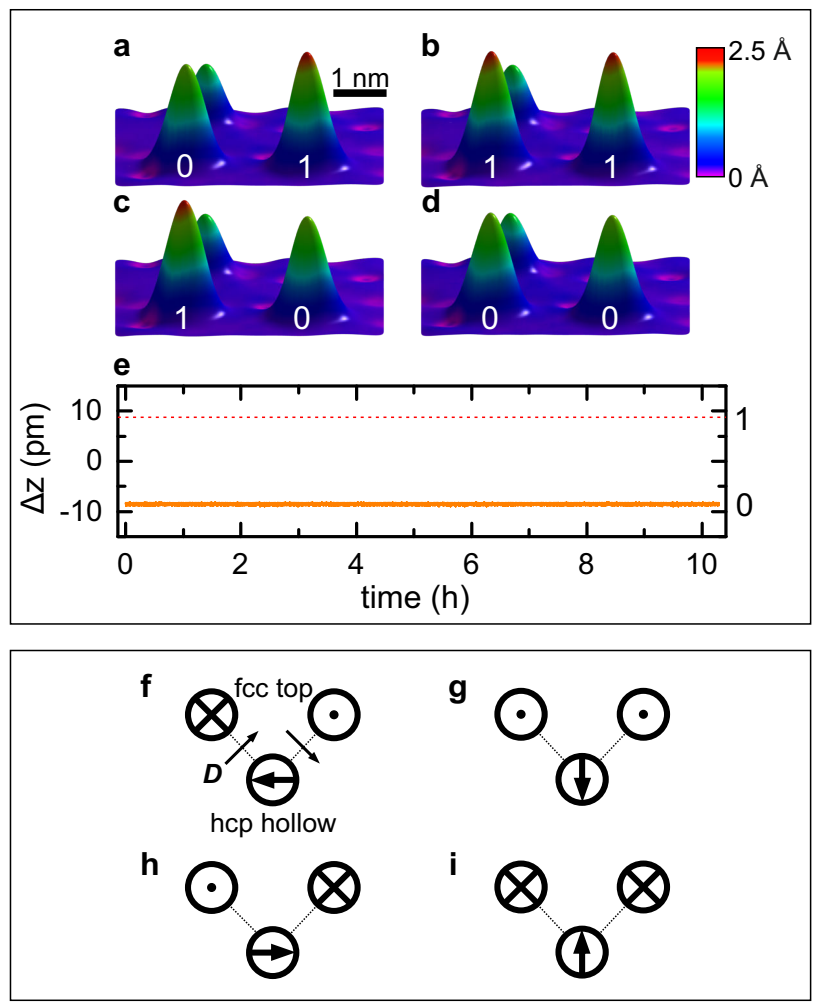

# SUPPLEMENTARY INFORMATION

**Supplementary Note 1** | **Determination of the GMC structure**

In order to determine the internal structure of the GMCs, we assembled several individual fcc and hcp atoms close to the built GMCs, as shown in the top of Supplementary Fig.1 exemplarily for one of the GMCs. The stacking type (fcc, hcp) of the individual Fe atoms is unambiguously identified by their characteristic ISTS signature [1]. Thereby, we can put a lattice of Pt surface atoms with assigned fcc and hcp adsorption sites on top of the GMCs topography (Supplementary Fig.1), and conclude that it contains either three fcc or three hcp atoms on nearest neighboring adsorption sites. Moreover, the GMC is not exactly round shaped, but slightly has the form of a downwards pointing triangle. We therefore conclude that this GMC is an fcc top cluster. The same procedure leads to the identification of the internal structure of the three other GMCs, which are shown schematically in the bottom of Supplementary Fig.1.

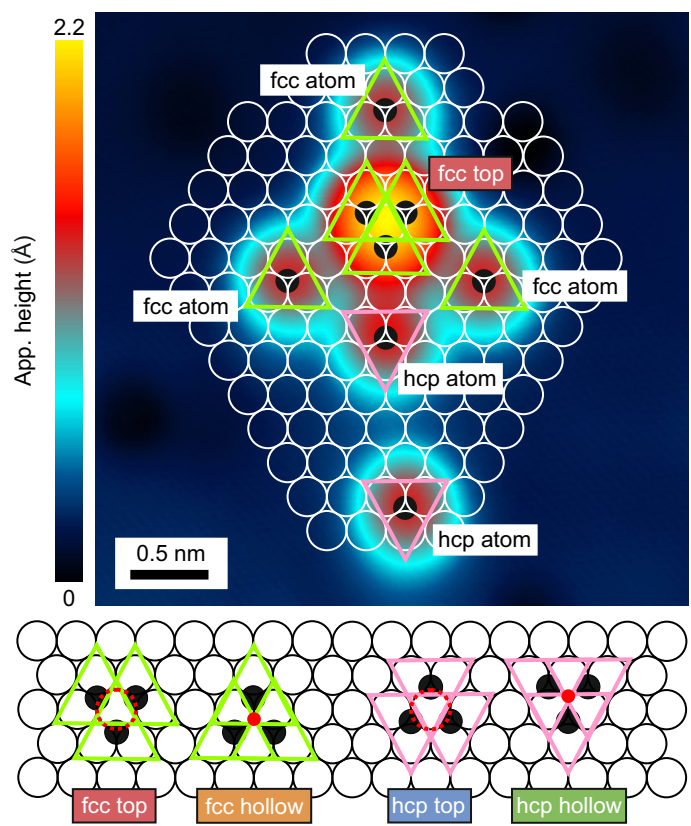

**Supplementary Figure 1 | GMC structure.** Constant-current image ($4 \times 4\,\mathrm{nm}^2$) of an assembly of a fcc top GMC, three individual fcc atoms, and two individual hcp atom close by. Tunneling parameters: $V = 5\,\mathrm{mV}$, $I = 500\,\mathrm{pA}$, $B = 0\,\mathrm{T}$, $T = 0.3\,\mathrm{K}$.

**Supplementary Note 2 | Density Functional Theory**

**Computational details** Density functional theory (DFT) calculations were performed in the framework of the Korringa-Kohn-Rostoker Green function (KKR-GF) real-space approach in the atomic sphere approximation with full charge density and including spin-orbit coupling [2,3]. The exchange and correlation effects were taken into account using the local spin-density approximation (LSDA) as parametrized by Vosko, Wilk and Nusair [4]. The Pt(111) surface was modeled by a slab of 24 Pt layers augmented by two vacuum regions 21.1 Å thick each, using the experimental lattice constant (a = 3.92 Å). To capture all the details of the hcp top and fcc top structures we have performed additional calculations using a slab containing 64 Pt atoms, corresponding to a 3x3 two-dimensional unit cell and 6 Pt layers (see next subsection and Supplementary Figure 2 for more details). The Fe trimers were constructed by cutting out a real space cluster centered on the positions to be occupied by the Fe atoms, which is embedded into the Pt substrate (see Fig. 2a of the main text). We have made sure that our results are converged with respect to the number of Pt atoms in the cluster, given that the MAE is found to be very sensitive to the Pt spin-polarization. Following this approach, we have modeled the four different structures analyzed in the main text, namely hcp hollow (194 Pt atoms in the cluster), hcp top (188 Pt atoms), fcc hollow (156 Pt atoms) and fcc top (170 Pt atoms).

For each of the four configurations, the vertical relaxation of the trimer was calculated using the QUANTUM-ESPRESSO package [5], imposing a relaxation criterion whereby the vertical force exerted on individual Fe atoms and Pt atoms of the surface layer is $<10^{-4}$ Ry a.u.$^{-1}$. The computational modeling of the system was done employing the repeated slab approach considering 86 atoms per unit cell, an energy cutoff of 40 Ry, a Gamma-point reciprocal-space mesh and ultrasoft fully relativistic pseudopotentials. In the hcp hollow and fcc hollow configurations, the trimer relaxes vertically 17.5% towards the surface (0% corresponds to the ideal interlayer separation in bulk, $a/\sqrt{3} = 2.26$ Å), and the height of the Pt surface layer remains approximately constant. In contrast, in the hcp top and fcc top configurations, one of the surface Pt atoms is shifted with respect to the rest of the Pt surface layer, as schematically illustrated in Supplementary Figure 2. Note that this particular Pt atom lies underneath the center of the Fe trimer and is therefore the only substrate atom that is a nearest neighbor of every Fe adatom. Finally, total energy calculations predict that the fcc and hcp hollow configurations are the energetically most stable ones (the calculated energy difference between the two is below our numerical precision), followed by fcc top (35 meV/adatom higher) and hcp top (45 meV/adatom higher).

**Magnetic moments, orbital moments and exchange parameters** The magnetic moments obtained using the KKR-GF approach are summarized in Supplementary Table 1. In all cases, we find that the magnetic moments of individual Fe adatoms are close to 3.5 $\mu_B$, while the surrounding Pt atoms have a substantial total contribution that ranges between 1.4 and 2.2 $\mu_B$ depending on the cluster type. It is particularly noteworthy that in the hcp top and fcc top configurations, the Pt atom with 10% vertical distance height (see previous subsection and Supplementary Figure 2) becomes strongly spin-polarized by 0.23 $\mu_B$, a factor 3 larger than its closest Pt neighbors. Finally,

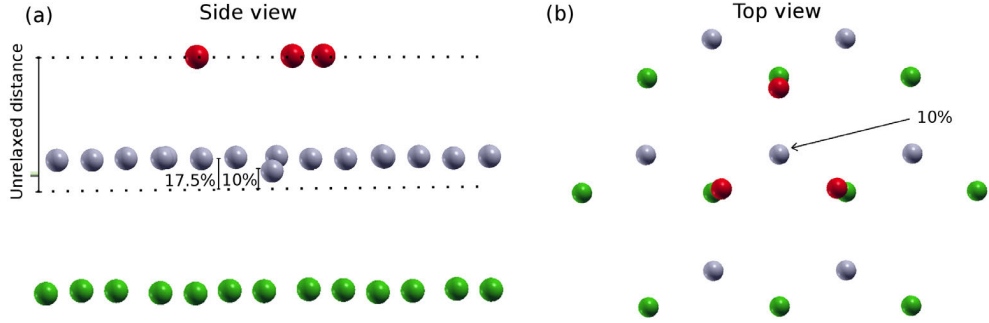

**Supplementary Figure 2 | Structure optimization.** (a) and (b) Respectively, side and top view of the relaxed structure of the hcp top cluster (only the trimer plus the first two Pt layers are shown for clarity). Fe adatoms, Pt surface atoms and Pt atoms from the second layer are symbolized by red, grey and green spheres, respectively. The surface Pt atom underneath the center of the Fe trimer relaxes to a different height as compared to the rest of the surface layer.

our calculations show a substantial orbital magnetic moment of both Fe adatoms and the Pt cluster.

The calculated magnetic exchange interaction parameters of the Fe trimer are summarized in Supplementary Table 2. For completeness, we have included calculations performed in the hcp top and fcc top structures where the single Pt surface atom underneath the center of the trimer was not shifted with respect to the rest of surface Pt atoms; we add a label 'c. h.' (constant height) to denote this structures. The MAE, denoted by $K$, has been calculated *ab initio* following the magnetic force theorem [6]. The rest of the parameters have been evaluated based on the generalized Lichtenstein formula [7–9]. Due to the $C_{3v}$ symmetry of the system the three Dzyaloshinskii–Moriya (DM) vectors are fully specified by two parameters, $D_\parallel$ and $D_\perp$ (see Suppl. Figs. 3 and 4):

$$\mathbf{D}_{12} = (0, -D_\parallel, D_\perp), \tag{1}$$

$$\mathbf{D}_{23} = (-\frac{\sqrt{3}}{2}D_\parallel, \frac{1}{2}D_\parallel, D_\perp), \tag{2}$$

$$\mathbf{D}_{31} = (\frac{\sqrt{3}}{2}D_\parallel, \frac{1}{2}D_\parallel, D_\perp). \tag{3}$$

|  | $m_{\text{sp., Fe}}$ ($\mu_B$) | $m_{\text{orb., Fe}}$ ($\mu_B$) | $m_{\text{sp., Pt cls.}}$ ($\mu_B$) | $m_{\text{orb., Pt cls.}}$ ($\mu_B$) |
|---|---|---|---|---|
| hcp hollow | 3.46 | 0.12 | 2.20 | 0.22 |
| hcp top | 3.33 | 0.12 | 2.03 | 0.08 |
| fcc hollow | 3.29 | 0.14 | 1.39 | 0.13 |
| fcc top | 3.28 | 0.12 | 1.72 | 0.12 |

**Supplementary Table 1 | Magnetic moments.** Calculated magnetic moments for the four different configurations considered in the main text. $m_{\text{sp., Fe}}$ and $m_{\text{orb., Fe}}$ denote respectively the magnetic and orbital moment per Fe adatom, while $m_{\text{sp., Pt cls.}}$ and $m_{\text{orb., Pt cls.}}$ refer to the magnetic and orbital moment of the full Pt cluster, respectively.

In addition, we have employed a scheme [8] that allows to incorporate the effect of the spin-polarized Pt substrate into the direct exchange coupling $J$, yielding a renormalized parameter $J_{\text{renorm}}$. We find that all the calculated $J$'s shown in Supplementary Table 2 are fairly large and negative, thus favouring a ferromagnetic coupling between the spins of the Fe trimer. Moreover, the renormalization induced by the Pt substrate has the net effect of increasing the magnitude of the $J$'s by approximately 20-30%, thus favouring even more the ferromagnetic coupling. In comparison, the DM interaction is nearly one order of magnitude smaller, except for the case of the hcp top structure, both in the normal and c. h. case. It is noteworthy that the $J/D_\parallel$ ratio varies by nearly 30%, thus revealing a large effect induced by the single shifted Pt atom.

**Non-collinear spin structure** Our DFT calculations have found a close competition between two almost ferromagnetic, slightly non-collinear spin configurations for the four trimers. In one of them, the magnetic moments point mainly along the out-of-plane direction with a non-collinearity polar angle $\theta$ as illustrated in Supplementary Figure 3. In the other one, shown in Supplementary Figure 4, the magnetic moments point nearly in-plane with two of them opening with a non-collinearity angle denoted as $x$. The corresponding band energy calculations, denoted by $\Delta E_b$, have been performed in two steps. Firstly, we have converged the solution to a strictly ferromagnetic out-of-plane configuration. Secondly, we have performed band energy calculations employing the magnetic force theorem [6] of canted configurations with varying opening angle, both for the setup illustrated in Supplementary Figures 3 and 4. The results are illustrated in Supplementary Figure 5 for all the configurations, including the 'c. h.' discussed in Supplementary Table 2. This figure shows that the nearly out-of-plane configuration is the most stable one for hcp top, fcc hollow and fcc top structures. In the particular case of hcp top, we note that it is necessary to consider the structure with the single shifted Pt atom in order to obtain the correct orientation of the magnetic moments that is in accordance with experimental measurements. Furthermore, we emphasize the importance of the energy minimization induced by the non-collinearity for the hcp top cluster; even though the ferromagnetic alignment favors the in-plane direction by nearly 1.5 meV/adatom, the energy gained by the non-collinearity reverses the trend and favours instead a nearly out-of-plane orientation with large canted angles of approximately $\theta = 17°$ by $\sim 0.1$ meV/adatom. In comparison, the fcc top and fcc hollow configurations show a small non-collinearity angle $\theta < 4°$. On the opposite side, the favoured configuration for the hcp hollow cluster type is the one with magnetic moments pointing nearly in-plane (see Supplementary Figure 4) with a non-collinearity angle of approximately $x = 4°$ (see Supplementary Figure 5).

|  | $K$ (meV/ad.) | $J$ (meV) | $J_{\text{renorm.}}$ (meV) | $D_\parallel$ (meV) | $D_\perp$ (meV) | non-col. angle (°) |
|---|---|---|---|---|---|---|
| hcp hollow | 0.65 | -52 | -71 | 8.7 | -5.0 | 4.1 |
| hcp top | 1.51 | -34 | -51 | 21.1 | -3.4 | 17.2 |
| hcp top c. h. | 1.23 | -34 | -47 | 10.7 | -4.4 | 6.5 |
| fcc hollow | -2.31 | -96 | -116 | 0.5 | -3.6 | $\sim 0$ |
| fcc top | -0.12 | -71 | -91 | 4.5 | -0.7 | 3.9 |
| fcc top c. h. | -0.21 | -82 | -108 | 2.5 | -1.3 | 1.5 |

**Supplementary Table 2 | Calculated exchange parameters.** The label 'c. h.' denotes the top configurations where the Pt surface atom underneath the center of the trimer was not shifted with respect to the rest of surface Pt atoms. The non-collinearity angle refers to the angle of the most stable non-collinear configuration, see Supplementary Figure 5.

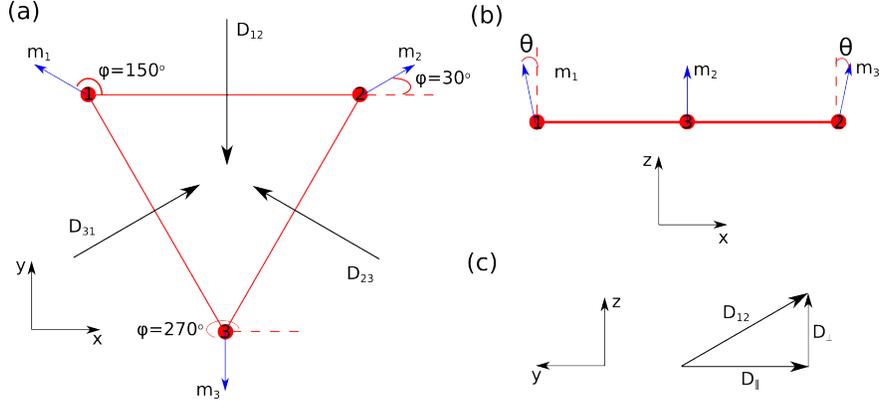

**Supplementary Figure 3 | Nearly out-of-plane spin configuration.** (a) and (b) respectively are top and side views of the Fe trimer in the nearly out-of-plane configuration. Red balls represent the adatoms with the label on top of the balls. Black arrows represent the DM vectors, while blue arrows represent the local magnetic moments (the in-plane projections of the magnetic moments have been enlarged for clarity). This spin structure respects the $C_{3v}$ symmetry of the system. (c) Definition of the parallel and perpendicular components of the DM vector.

The opening of the non-collinearity angles can be related to the exchange parameters by making use of a classical Heisenberg model:

$$H = +J \sum_{i<j} \hat{\mathbf{m}}_i \cdot \hat{\mathbf{m}}_j + \sum_{i<j} \mathbf{D}_{ij} \cdot (\hat{\mathbf{m}}_i \times \hat{\mathbf{m}}_j) + K \sum_i m_{i,z}^2 \equiv H_J + H_D + H_K, \qquad (4)$$

where $\hat{\mathbf{m}}_i = (\cos\varphi_i \sin\theta_i, \sin\varphi_i \sin\theta, \cos\theta_i), \ i = 1, 2, 3$ are the magnetization unit vectors of the adatoms arranged as in Supplementary Figures 3 and 4. Note that within the present convention $J < 0$ ($J > 0$) gives rise to a ferromagnetic (antiferromagnetic) coupling, and $K < 0$ ($K > 0$) favors out-of-plane (in-plane) orientation of the magnetic moments.

We now compute an analytic solution for the direction of the spin moments illustrated in Supplementary Figures 3 and 4. Let us begin with the nearly in-plane configuration and thus set the azimuthal angle of all adatoms to $\varphi_i = 0$, *i.e.* the direction of the spin-moments are

$$\hat{\mathbf{m}}_i = (\sin\theta_i, 0, \cos\theta_i), \ i = 1, 2, 3. \qquad (5)$$

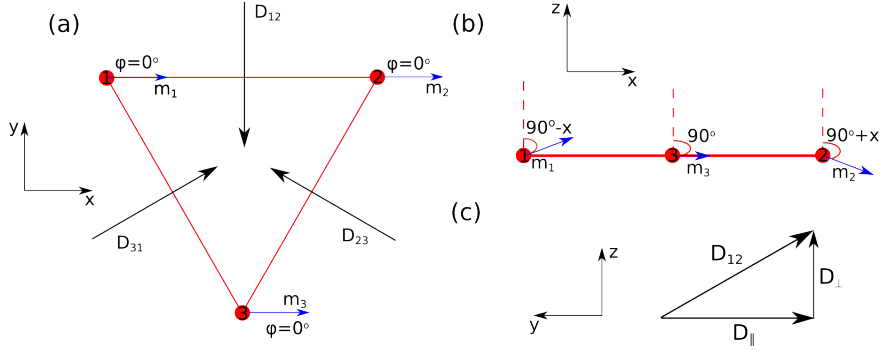

**Supplementary Figure 4 | Nearly in-plane spin configuration.** (a) and (b) respectively are top and side views of the Fe trimer in the nearly in-plane configuration. Labels are as in Supplementary Figure 3. We note that, although this spin structure breaks the $C_{3v}$ symmetry of the system, there are two other solutions which are degenerate in energy with it and transform into each other according to the $C_{3v}$ symmetry. (c) Definition of the parallel and perpendicular components of the DM vector.

The three contributions to the Hamiltonian of Eq. 4 are then given by:

$$H_J = J \sum_{i<j=1}^{3} \left( \sin\theta_i \sin\theta_j + \cos\theta_i \cos\theta_j \right), \tag{6}$$

$$H_D = -D_\| \left( \cos\theta_1 \sin\theta_2 - \cos\theta_2 \sin\theta_1 \right)$$
$$+ \frac{1}{2} D_\| \left( \cos\theta_2 \sin\theta_3 - \cos\theta_3 \sin\theta_2 \right) + \frac{1}{2} D_\| \left( \cos\theta_3 \sin\theta_1 - \cos\theta_1 \sin\theta_3 \right), \tag{7}$$

$$H_K = K \left( \cos^2\theta_1 + \cos^2\theta_2 + \cos^2\theta_3 \right). \tag{8}$$

We seek next for solutions near $\theta_i = \pi/2$ by using $\sin\theta_i = \sin(\pi/2 + x_i) = \cos x_i \sim 1 - x_i^2/2$, $\cos\theta_i = \cos(\pi/2 + x_i) = -\sin x_i \sim -x_i$, allowing to write the full Hamiltonian as

$$H = J\left(x_1 x_2 + x_1 x_3 + x_2 x_3\right) + K\left(x_1^2 + x_2^2 + x_3^2\right) + \frac{3}{2} D_\|\left(x_1 - x_2\right) + 3J. \tag{9}$$

The $x_i$ that minimize this Hamiltonian are obtained from the solution of three coupled equations $\partial H/\partial x_i = 0$, yielding

$$\begin{aligned} x_1 &= +\frac{3}{2} \frac{D_\|}{J - 2K}, \\ x_2 &= -\frac{3}{2} \frac{D_\|}{J - 2K}, \\ x_3 &= 0. \end{aligned} \tag{10}$$

The corresponding total energy for the nearly in-plane configuration is then given by

$$E_{\text{in}} \equiv \frac{9 D_\|^2}{4J - 8K} + 3J \sim \frac{5 D_\|^2}{4J} + 3J. \tag{11}$$

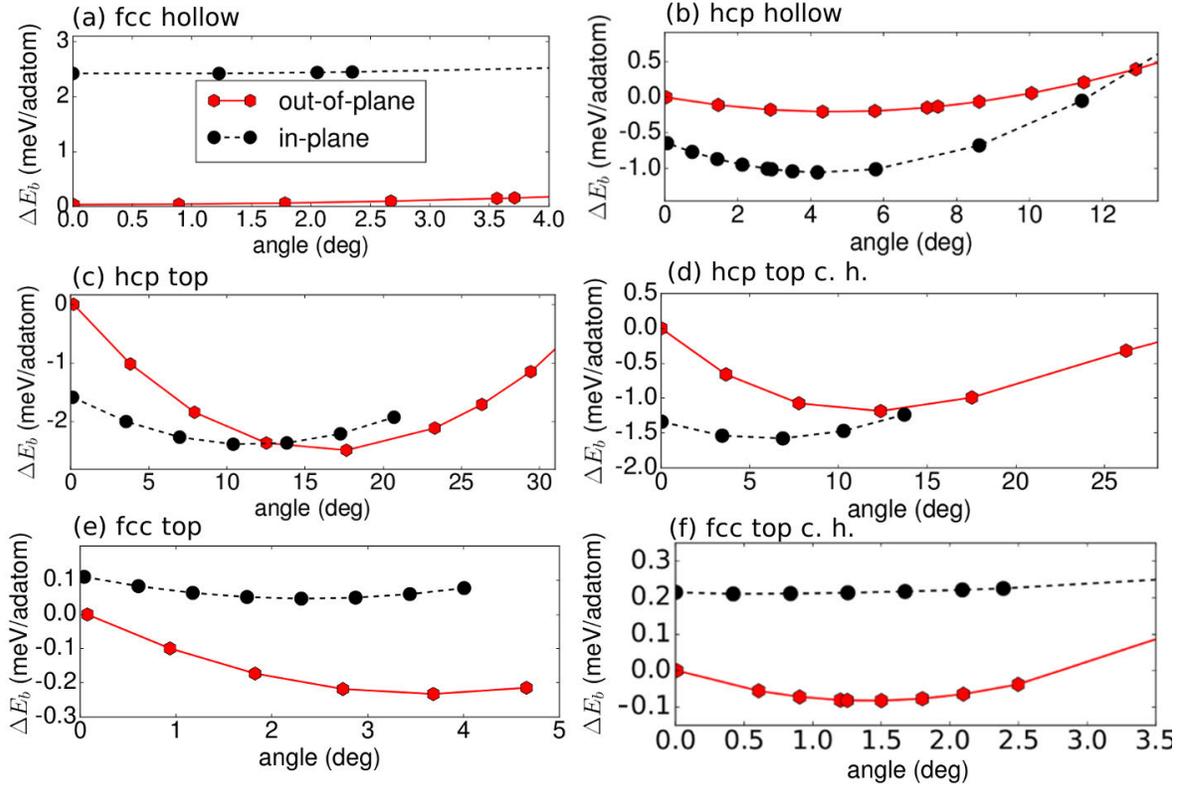

**Supplementary Figure 5 | Non-collinear band calculations.** DFT non-collinear band-energy calculations following the magnetic force theorem. Red and black points denote the spin structure depicted in Supplementary Figures 3 and 4, respectively.

Given that $J$ is negative (ferromagnetic coupling), this energy is also negative.

We next turn to the nearly out-of-plane configuration illustrated in Supplementary Figure 3 by fixing the azimuthal angles $\varphi_1 = 30°$, $\varphi_2 = 150°$ and $\varphi_3 = 270°$ and setting the same polar angle for the three adatoms:

$$\hat{\mathbf{m}}_1 = (-\frac{\sqrt{3}}{2}\sin\theta, \frac{1}{2}\sin\theta, \cos\theta), \tag{12}$$

$$\hat{\mathbf{m}}_2 = (\frac{\sqrt{3}}{2}\sin\theta, \frac{1}{2}\sin\theta, \cos\theta), \tag{13}$$

$$\hat{\mathbf{m}}_3 = (0, -\sin\theta, \cos\theta). \tag{14}$$

Then, the three parts of the Hamiltonian can be expressed as (after straightforward algebra in the $J$ and DM parts):

$$H_J = 3J\left(-\frac{1}{2}\sin^2\theta + \cos^2\theta\right), \tag{15}$$

$$H_D = 3\sqrt{3}D_\parallel \sin\theta\cos\theta - \frac{3\sqrt{3}}{2}D_\perp \sin^2\theta, \tag{16}$$

$$H_K = 3K\cos^2\theta. \tag{17}$$

We next assume small deviations $\sin\theta \sim \theta$, $\cos\theta \sim 1-\theta^2/2$, allowing to write the full Hamiltonian as

$$H = -\frac{3}{2}J\theta^2 + 3\sqrt{3}D_\parallel x - \frac{3\sqrt{3}}{2}D_\perp \theta^2 - 3K\theta^2 + 3K + 3J. \tag{18}$$

The solution of $\partial H/\partial\theta = 0$ is

$$\theta = \frac{\sqrt{3}D_\parallel}{J + 2K + \sqrt{3}D_\perp}. \tag{19}$$

In all cluster types studied in this work the hierarchy $|J| \gg |D_\parallel|, |D_\perp|, |K|$ is fulfilled. Therefore, the angles predicted by Eqs. 10 and 19 reveal a larger impact of non-collinearity on the nearly out-of-plane configuration than in the nearly in-plane configuration by a factor of approximately $2/\sqrt{3} \sim 1.15$, which is in reasonable accordance with the *ab initio* band calculations shown in Supplementary Figure 5.

Inserting Eq. 19 into Eq. 18 one obtains the total minimum energy of the nearly out-of-plane configuration,

$$E_{\text{out}} = 3(K+J) + \frac{9}{2}\frac{D_\parallel^2}{\sqrt{3}D_\perp + J + 2K} \sim \frac{9D_\parallel^2}{2J} + 3K + 3J. \tag{20}$$

Thus, the energy difference between Eq. 11 and Eq. 20 gives the relative energy gain induced by non-collinearity between the nearly in-plane and out-of-plane configurations:

$$E_{\text{out}} - E_{\text{in}} \sim \frac{9D_\parallel^2}{4J} + 3K. \tag{21}$$

The above equation shows that, in absence of MAE ($K \sim 0$), $E_{\text{out}} - E_{\text{in}} < 0$ as long as the coupling is ferromagnetic, *i.e.* the nearly out-of-plane direction is favoured by non-collinearity. Interestingly, even if $K > 0$ (MAE favours in-plane orientation), the overal total energy may still favour the nearly out-of-plane configuration provided $\left|9D_\parallel^2/4J\right| > |3K|$ is fulfilled.

## Supplementary Note 3 | Fano functions

In order to extract the magnetic field dependent shift and temperature induced broadening of the Kondo resonance (Fig.3 of the main manuscript), we fitted the measured voltage dependent spectra $\frac{dI}{dV}(V)$ to the following sum of two Fano functions [10] including a linear background:

$$G(V) = G_1 \cdot \frac{(q+\varepsilon_1)^2}{1+\varepsilon_1^2} + G_2 \cdot \frac{(q+\varepsilon_2)^2}{1+\varepsilon_2^2} + G_{\text{off}} + G_{\text{lin}} \cdot V \qquad (22)$$

$$\varepsilon_i = \frac{eV - E_i}{\Gamma} \qquad (23)$$

Here, $q$ is the so-called form factor, $\Gamma$ is the full width at half maximum and $E_i$ are the energetic positions of the Kondo resonances. The parameters extracted by fitting such functions to the experimental data are given in Supplementary Tables 3 and 4, and the two fitted Fano functions for the case of the magnetic field dependence are shown in Supplementary Fig.6. Note, that the resulting $q$-factors are negative, indicating strong phase shifts between the two tunneling paths [11].

**Supplementary Table 3 | Parameters for magnetic field dependent Fano functions.** Parameters used for the fits of the $B$ dependent spectra of the hcp-hollow cluster (Fig.3a of main manuscript).

| $B$ (T) | $\Gamma$ (meV) | $E_1$ (meV) | $E_2$ (meV) | $q$ | $G_1$ (a.u.) | $G_2$ (a.u.) |
|---|---|---|---|---|---|---|
| 0 | 0.32 | -0.05 | - | -1.60 | 1.68 | - |
| 1.5 | 0.42 | 0.15 | -0.14 | -1.52 | 1.99 | 1.44 |
| 3 | 0.34 | 0.25 | -0.43 | -2.98 | 0.73 | 0.49 |
| 4.5 | 0.37 | 0.31 | -0.60 | -5.01 | 0.31 | 0.18 |
| 6 | 0.53 | 0.49 | -0.59 | -3.08 | 0.79 | 0.26 |
| 7.5 | 0.73 | 0.69 | -0.80 | -2.30 | 1.36 | 0.19 |

**Supplementary Table 4 | Parameters for temperature dependent Fano functions.** Parameters used for the fits of the $T$ dependent spectra measured on the hcp-hollow cluster (Fig.3b of main manuscript).

| $T$ (K) | $\Gamma$ (meV) | $E_1$ (meV) | $q$ | $G_1$ (a.u.) |
|---|---|---|---|---|
| 0.31 | 0.40 | 0.054 | -0.78 | 5.3 |
| 1.3 | 0.70 | 0.062 | -0.81 | 4.3 |
| 2.3 | 1.30 | 0.031 | -0.89 | 3.9 |
| 3.1 | 1.62 | 0.022 | -0.80 | 4.1 |
| 4.3 | 2.13 | -0.034 | -1.04 | 3.3 |
| 5.3 | 2.82 | -0.126 | -1.04 | 3.8 |
| 6.2 | 3.10 | 0.058 | -1.22 | 2.7 |

In order to estimate the Kondo temperature $T_K$, the resulting temperature dependence of $\Gamma$ was fitted to the power law [12] $\Gamma(T) = \frac{1}{2}\sqrt{(\alpha k_B T)^2 + (2k_B T_K)^2}$ resulting in $T_K = 4.19\,\text{K}$ ($\alpha = 11.8$) which is illustrated as a grey line in Fig.3d together with the data. Additionally, we compare to published numerical renormalization group calculations for a spin-1/2 impurity in the strong coupling regime [13], given by the grey dots in Fig.3d, resulting in a similar value of $T_K = 4.64\,\text{K}$.

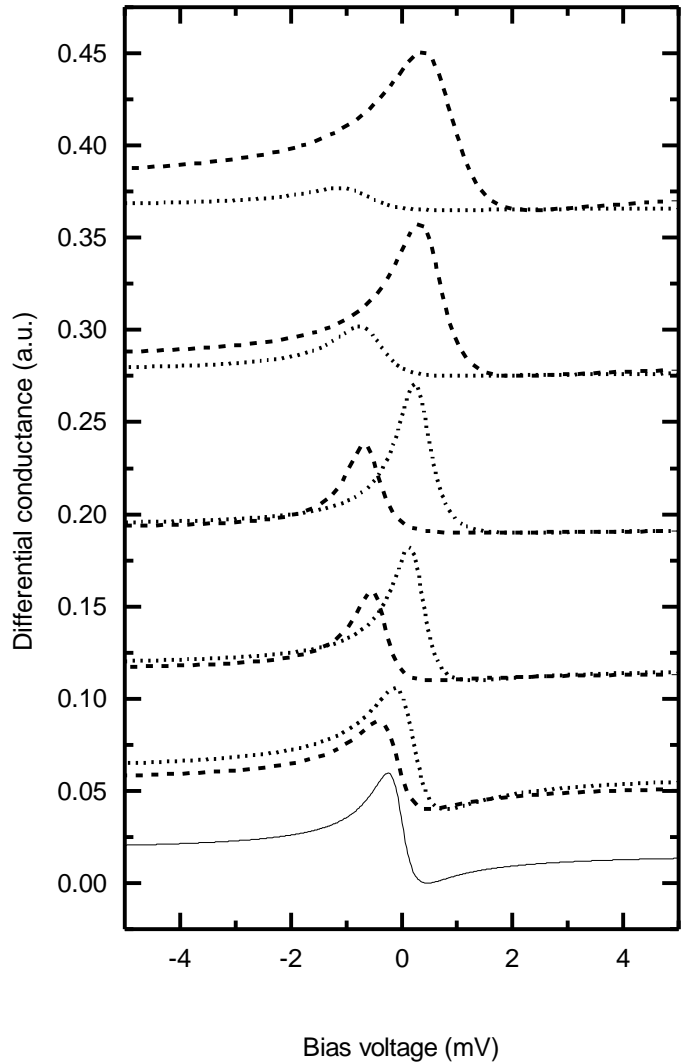

**Supplementary Figure 6 | Fano functions.** Here, the used Fano functions are plotted without the linear background ($G_{\text{lin}} = 0$) but with artificial offsets for better visibility.

## Supplementary Note 4 | ISTS

Supplementary Fig.7 (left panels) show $dI/dV$ curves measured on the hcp top, fcc hollow and fcc top GMCs as a function of the out-of-plane oriented magnetic field. They reveal features at positive and negative bias voltage linearly increasing symmetrically around the Fermi energy ($V = 0\,\text{mV}$) as a function of $B$. The associated $d^2I/dV^2$ curves (right panels) have peaks and dips at the corresponding bias voltages. This behavior is the fingerprint of a spin excitation of an out-of-plane easy axis system [1]. Note, however, that the shape of the features in the spin excitation spectra partly deviates from the usual step-like appearance typically found for magnetic atoms which are more strongly decoupled from a metallic substrate [14,15]. This might indicate deviations from the simple effective-spin model where the magnetic impurity is artificially separated into an interior part (effective spin) that is excited by the tunneling electrons, and an exterior part that interacts with the effective spin leading to damping of the spin excitations. In the present system, this border is not well defined due to the GMC character of the magnetic impurity, which can lead to excitation spectra that strongly deviate from the simple step shape [16]. We extract the corresponding excitation energies $\Delta_{01}$ by searching the center of the symmetric peaks/dips in the $d^2I/dV^2$ curves, which are marked by crosses, both in the $dI/dV$ as well as in the $d^2I/dV^2$ curves. The extracted excitation energies $\Delta_{01}$ are plotted in Supplementary Fig.8 as a function of $B$. Indeed, there is a linear behavior, as expected from the out-of-plane system. By fitting a linear function to these plots, we extracted the $g$-factors via $g = 1/\mu_\text{B} \cdot d\Delta_{01}(B)/dB$. The accordingly determined parameters $\Delta_{01}$ and $g$ of the three GMCs are given in Supplementary Table 5. Here, we distinguish between the values determined from the excitation on the negative (neg.) and positive (pos.) bias voltage sides, which allows us to estimate the error, which is given in the Table of the main manuscript text.

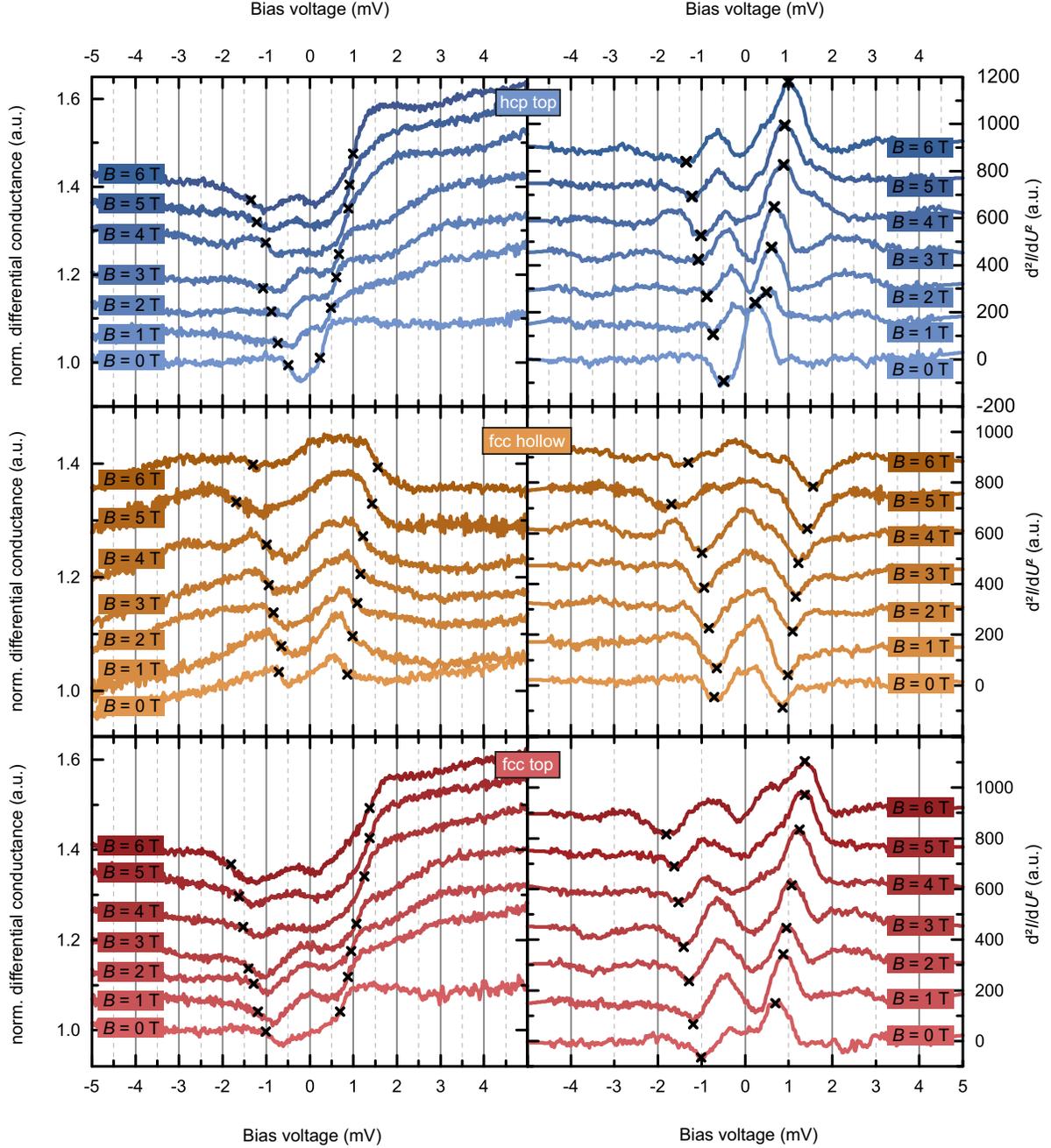

**Supplementary Figure 7 | ISTS of hcp top, fcc hollow and fcc top GMCs.** $dI/dV$ (left panels) and $d^2I/dV^2$ (right panels) as calculated by smoothing (every datapoint is the average of the 50 surrounding datapoints) and consecutive numerical differentiation of the $dI/dV$ spectra that have been measured by Lock-In technique. The crosses mark the positions of the extracted spin-excitation energies $\Delta_{01}$. Tunneling parameters: $V_{\text{stab}} = 5\,\text{mV}$, $I_{\text{stab}} = 2\,\text{nA}$, $V_{\text{mod}} = 80\,\mu\text{V}$, $T = 0.3\,\text{K}$.

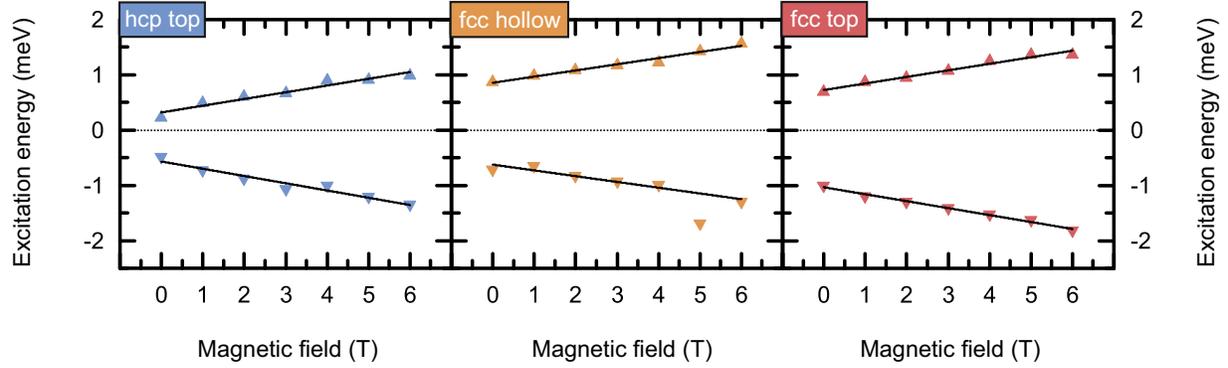

**Supplementary Figure 8 | Spin excitation energies determined from ISTS.** The plotted spin excitation energies $\Delta_{01}$ (markers) were determined from the spectra in Supplementary Fig.7. The lines are fits of the excitation energies on the positive and negative side to linear functions.

**Supplementary Table 5 | Zero field excitation energies $\Delta_{01}$ and $g$-factors.** The values were determined from the zero field excitation energies and fitted lines on the negative (neg.) and positive (pos.) energy side in Supplementary Fig.8.

| cluster type | $\Delta_{01}$ (meV, neg.) | $\Delta_{01}$ (meV, pos.) | $g$ (neg.) | $g$ (pos.) |
| --- | --- | --- | --- | --- |
| hcp top | -0.57 | 0.32 | 2.26 | 2.11 |
| fcc hollow | -0.62 | 0.86 | 1.8 | 1.92 |
| fcc top | -1.03 | 0.73 | 2.18 | 2.06 |

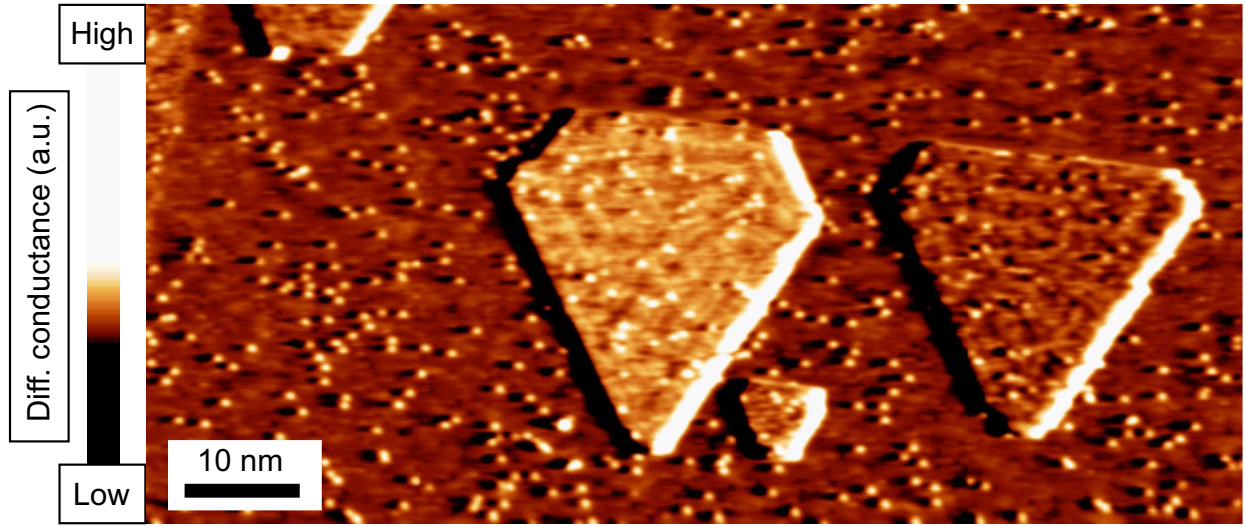

**Supplementary Figure 9 | Spectroscopic SPSTM image of Cobalt islands on Pt(111).** Differential conductance image of the sample with triangular shaped Co monolayer islands and Fe atoms on Pt(111). The large left (right) islands appears bright (dark) indicating their opposite out-of-plane magnetizations and the sensitivity of the tip to the out-of-plane component of the sample magnetization. Tunneling parameters: $V = -10\,\mathrm{mV}$, $I = 750\,\mathrm{pA}$, $V_\mathrm{mod} = 5\,\mathrm{mV}$, $B = 0\,\mathrm{T}$, $T = 0.3\,\mathrm{K}$.